\documentclass[twocolumn,aps,tightenlines,floatfix,showpacs]{revtex4}
\usepackage{graphicx}
\begin{document}

\title{$  T=1  $ Pairing Along the $  N=Z  $ Line}

\author{B. Alex Brown}

\affiliation{The Facility for Rare Isotope Beams and Department
of Physics and Astronomy, Michigan state University, East Lansing,
Michigan 48824-1321, USA}

\author{Michio Honma}

\affiliation{Center for Mathematical Sciences, University of Aizu, Tsuruga,
Ikki-machi, Aizu-Wakamatsu, Fukushima 965-8580, Japan}

\author{Ragnar Stroberg}

\affiliation{Department of Physics and Astronomy, University of
Notre Dame, Notre Dame, IN, 46556, USA}

\begin{abstract}
Pairing energies for the addition of two neutrons
on even-even nuclei with $  N=Z  $ are studied. The $  Z  $
dependence is attributed to the number and type
of orbitals that are occupied in the valence shell-model space.
Properties in the region from $  Z=60-100  $ depend on the
location of the $  0g_{9/2}  $ orbital.

\end{abstract}

\maketitle

Nuclear masses in the region of $  A=70  $ were measured and
recently reported by Wang et al. \cite{wang}. These results were used to
extract the quantity $  \delta V^{oo}_{pn}  $ for odd-odd nuclei
defined in terms of the nuclear binding energies $  B  $ by
the double-difference equation
related to the arrows (a) and (b) in Fig. 1:
$$
\delta V^{oo}_{pn} = \Delta B_{a} - \Delta B_{b},       \eqno({1})
$$
where
$$
\Delta B_{a} = B(Z,N) - B(Z,N-1)       \eqno({2})
$$
and
$$
\Delta B_{b} = B(Z-1,N) - B(Z-1,N-1).       \eqno({3})
$$
The cases of interest involve an even-even core nucleus,
$  A_{c}  $, with $  T=0  $ and $  N_{c}=Z_{c}  $, and the addition of a proton
and neutron coupled to $  T=1  $.
It was shown that the experimental values for this quantity
were much higher than all of the theoretical calculations for
$^{66}$Ga, $^{70}$Br and $^{74}$Rb used for Fig. 2 of \cite{wang}.
The reason for this difference was investigated using
valence-space in-medium similarity renormalization
group (VS-IMSRG)
method in the $  fp  $ $  (0f_{7/2},0f_{5/2},1p_{3/2},1p_{1/2})  $ model space.
The purpose of this paper it to put these results in the
context of other data and calculations along the $  N=Z  $ line of nuclei.
\begin{figure}
\includegraphics[scale=0.5]{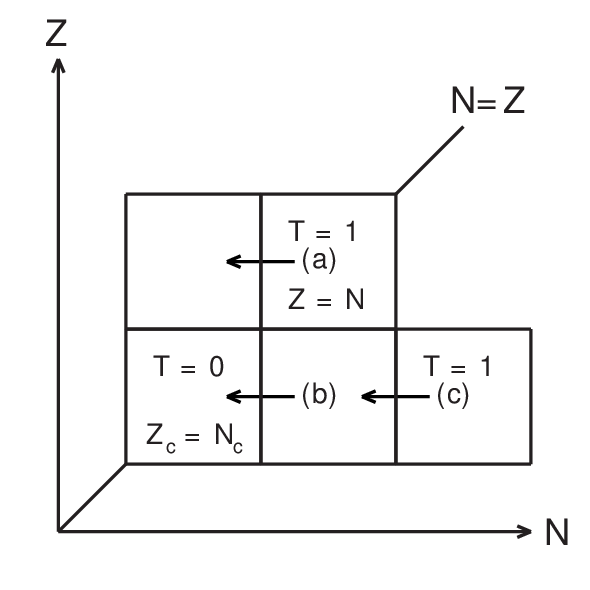}
\caption{Segment of the nuclear chart near $  N=Z  $
showing the binding-energy differences discussed in the
text.
}
\end{figure}

The series of odd-odd nuclei $^{62}$Ga, $^{66}$As, $^{70}$Br and $^{74}$Rb
used for Fig. 2 of \cite{wang} all have $  J^{\pi }  $=0$^{ + }$, $  T=1  $
ground states.
For these cases, the ground states of the nuclei connected by
arrow (c) in Fig. 1 are isobaric analogues
of the ground states connected by arrow (a).
Thus, results for $  T=1  $ pairing can also be obtained using
the binding-energy differences for the nuclei
shown by the arrows (c) and (b) in Fig. 1:
$$
D_{n}(Z-1,N) = \Delta B_{c} - \Delta B_{b},       \eqno({4})
$$
where
$$
\Delta B_{c} = B(Z-1,N+1) - B(Z-1,N).       \eqno({5})
$$

This is the pairing-gap equation defined in \cite{low}.
For example, for $^{74}$Rb, we have from Eq. (1),
$  \delta V^{oo}_{pn}  $ = 13.976(40) - 10.682(11) = 3.294(42) MeV,
and from Eq. (4), $  D_{n}  $ = 13.851(7) - 10.682(11) = 3.169(13) MeV.
The difference between these two results is due to the
small charge-symmetry breaking interaction in the $  T=1  $
triplet, as well as the small $  N  $
dependence in the binding-energy differences between isobaric analogue
states.

In Fig. 2a  we show $  D_{n}  $ and $  \delta V_{pn}(T=1)  $ for the even-even
core nuclei with $  N_{c}=Z_{c}  $
from $  Z_{c}=4  $ up to   $  Z_{c}=50  $. The binding energies are
from the 2020 mass table \cite{mass} together with the
new data from \cite{wang}. For the odd-odd nuclei involved in
$  \delta V_{pn}(T=1)  $ we use the binding energy associated with the
0$^{ + }$, $  T=1  $ states which in some cases are excited states.
Binding energy for nuclei
near $  N=Z  $ above about $  A=76  $
are not measured, but based on mass-value extrapolations
with relatively large uncertainties.
The region of data considered in \cite{wang} is $  A_{c}=56-72  $
in Fig. 2. The largest difference between $  D_{n}  $ and
$  \delta V_{pn}(T=1)  $  is for $  A_{c}=68  $. The masses involved
in the results for $  A_{c}=68  $ should be
experimentally confirmed.

\begin{figure}
\includegraphics[scale=0.5]{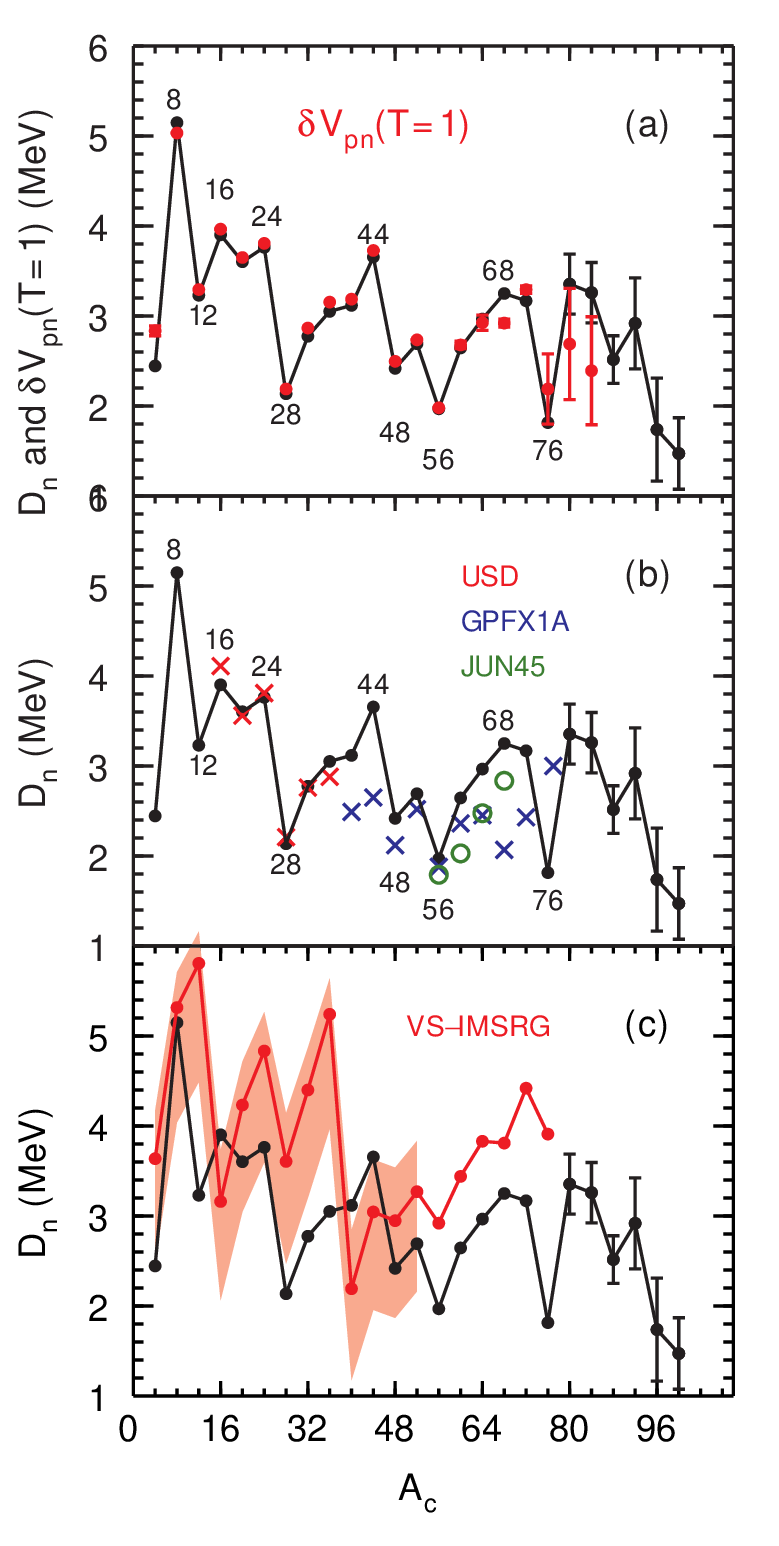}
\caption{
 $  D_{n}  $ for the even-even
core-nuclei with $  N_{c}=Z_{c}  $from $  Z_{c}=2  $ up to   $  Z_{c}=50  $.
The experimental data are shown
by the black cicles with error bars (same for all panels).
In panel (a) the data for $  D_{n}  $ are compared to the data for
$  \delta V_{pn}(T=1)  $ as obtained from Eq. (1).
In panel (b) the data for $  D_{n}  $
are compared to those obtained with empirical
shell-model Hamiltonians in various model space.
The red crosses are the
results of calculations in the $  ds  $ model space
with the USDC Hamiltonian \cite{usdc}.
 The blue crosses are the
results of calculations in the $  fp  $ model space
with the GPFX1A Hamiltonian \cite{gx1a}.
 The green circles are the
results of calculations in the $  jj44  $ model space
with the JUN45 Hamiltonian \cite{jun45}.
In panel (c) the data for $  D_{n}  $
are compared to the results of VS-IMSRG
calculations shown in red.
}
\end{figure}

In Fig. 2b the experimental data for $  D_{n}  $ are
compared with the results of configuration-interaction calculations.
The results for $  A_{c}=16-36  $ were obtained
in the $  ds  $ $  (0d_{5/2},0d_{3/2},1s_{1/2})  $
model space with the USDC Hamiltonians \cite{usdc} are shown by the
red crosses in Fig. 2b. The agreement is excellent.
The same level of agreement would
be obtained with the earlier versions of the "universal" $  ds  $
Hamiltonians
USD \cite{usd}, USDA \cite{usda} and USDB \cite{usda}.
This agreement it not surprising since these are empirical
Hamiltonians whose two-body matrix elements (TBME) are obtained
by singular-valued decomposition fits to binding energies
and excitation energy data for nuclei in
the region of $  Z=8-20  $ and $  N=8-20  $. All energy data in the $  ds  $
model space can be
described by a unified set of TBME within an rms uncertainty
of about 150 keV, except for those in the region
around $^{32}$Mg that lie within the island-or-inversion \cite{ioi},
where one or more neutrons are in the $  fp  $ shell
in the ground states.

The results for $  A_{c}=40-78  $ were obtained in the $  fp  $
model space with the GPFX1A Hamiltonians \cite{gx1a} are shown by the
blue crosses in Fig. 2b.
Binding energies and excitation energies for nuclei
in the region $  A \geq 47  $ and $  Z \leq 32  $ were used
to determined the universal effective TBME for GPFX1A.  It is not
suprising that the calculated $  D_{n}  $ values are in
relatively good agreement with experiment for this region.
Data for $  A=40-46  $ was
not included since the energies are
influenced by mixing with low-lying intruder states
coming from particle-hole states across $  Z=20  $
and $  N=20  $.
The calculated pairing for $  A=40  $ and $  A=44  $ are significantly
smaller than experiment. At  $  A=60  $ and above theory and experiment
start to diverge.

The irregular patterns found in the experimental data
are related to the types of valence orbitals
involved in the configurations. Just below
$  jj  $ magic numbers $  A_{c}  $=12, 28 and 56,
the pairing is dominated by high $  j  $
orbitals  ($  0p_{3/2}  $, $  0d_{5/2}  $
and $  0f_{7/2}  $, respectively) giving a relatively large
pairing energy. Just after these magic number a lower $  j  $
orbital starts to be filled
  ($  0p_{1/2}  $, $  1s_{1/2}  $
and $  1p_{3/2}  $, respectively) and the pairing energy drops.

For $  A_{c} \geq 64  $, the $  0g_{9/2}  $ orbital
becomes important. This can be seen in the
spectra of nuclei with $  A_{c}+1  $. Energies
of the lowest 9/2$^{ + }$ states are  2.40 MeV ($^{61}$Zn),
1.22 MeV ($^{65}$Ge), 0.57 MeV ($^{69}$Se)
and 0.43 MeV ($^{73}$Kr). Starting in $^{77}$Sr
the nuclei are more deformed and the ground
state has 5/2$^{ + }$. Calculations for pairing in this
region must explicitly take into account
the growing importance of the $  0g_{9/2}  $
orbital. The $  fp  $ model space used
for the interpretation of the data
in \cite{wang} is insufficient.

In a similar way, pairing with the  $  0f_{7/2}  $
orbital in the upper part of the $  ds  $
model space must influence the $  D_{n}  $ values.
7/2$^{-}$ states appear at 3.62 MeV ($^{29}$Si),
2.93 MeV ($^{33}$S) and 1.61 MeV ($^{37}$Ar).
In contrast to $  0g_{9/2}  $ in the $  A=70  $ region
mixing with $  0f_{7/2}  $ in the upper $  ds  $ shell
appears to be small enough to be contained implicitly
as a perturbative contribution to the $  ds  $
effective TBME.
In contrast, in the $  A=70  $ region, the $  0g_{9/2}  $
must be treated explicitly in the pairing.
Then one expects $  D_{n}  $ to become larger
than those obtained in the $  fp  $ model space due to mixing with
this high-$  j  $ orbital.
An interesting feature for $  A_{c}>60  $ in Fig. 2 is the sharp drop
at $  A_{c}=76  $. Perhaps this is due to a sudden shape change.

In order to explain the data, the calculations presented in \cite{wang}
used the valence-space in-medium similarity renormalization
group (VS-IMSRG)
method with and without the 3N force from the chiral EFT interaction
in the
$  fp  $ model space. To put this comparison in a broader
context, we show in Fig. 2c
results based on binding energies obtained with VS-IMSRG calculations
as given in the supplemental material obtained of \cite{vs}, for
the  $  0p  $ ($  A_{c}  $=8,12), $  ds  $ ($  A_{c}  $=16-36) and
$  fp  $ ($  A_{c}  $=40-52) model spaces.
The error band takes into account
the estimated 0.8 MeV error in the one-neutron separation energies found in
\cite{vs}. The error band also includes an estimated systematic downward
shift of 0.2 to 0.4 MeV that takes into account that the interaction gives rms radii
that are systematically too small.
For $  A_{c}  $=56-76 we show the results of VS-IMSRG calculations
using the same method as \cite{vs}.
The error band for  $  A_{c} \geq 56  $ has
not been evaluated, but it should be similar those shown for  $  A_{c} \leq 52  $.
The overall size of the
VS-IMSRG results for $  D_{n}  $ are qualitatively consistent
with experiment within the error band, but they are generally
larger than experiment.
In order to draw conclusions
about the deficiencies of the VS-IMSRG results for $  A_{c} \geq 64  $,
one must first understand and improve the results for $  A_{c} \leq 64  $.

The minimal model space appropriate for the region above $  A_{c}=60  $
must involve the orbitals $  (0f_{5/2},1p_{3/2},1p_{1/2},0g_{9/2})  $
(the $  jj44  $ model space) \cite{jun45}, \cite{jj44}.
Calculations within the $  jj44  $ model space must be able to
reproduce the low-lying energy of the $  0g_{9/2}  $ orbital
and the shape change around $^{80}$Zr. The full deformation
of $^{80}$Zr also requires the addition of the $  1d_{5/2}  $ orbital \cite{d5}.

For $  A_{c}  $=56, 60, 64, and 66 we show the results obtained with the JUN45
hamiltonian in the $  jj44  $ model space. For $  A_{c}  $=68
the $  jj44  $ $  D_{n}  $ value is about one MeV  larger than the $  fp  $ model
space result. This is correlated with a change
in the occupation of the $  0g_{9/2}  $ from 0.48 in $^{66}$Ge
to 1.48 in $^{70}$Se.
The basis dimensions in the $  jj44  $
model space for $  A_{c}\geq 70  $ are very large,
and the consistent set of calculations required
to reproduce the spectroscopy in this region
is beyond the scope of this short paper.

In summary, we have used experimental data on binding energies
to deduce the $  T=1  $ pairing energy related to the addition
of two neutrons in nuclei with $  N=Z  $ from $^{8}$Be to $^{100}$Sn. We observe
trends that are correlated with the change of the orbitals that are
occupied along this set of nuclei. In the regions of
nuclei described by the $  ds  $ and $  fp  $ model spaces,
the  "universal" effective Hamiltonians describe the data, as expected,
since they are obtained from fits to energy data in these model spaces.
The results obtained from the VS-IMSRG method in these same model
spaces are not good enough to draw conclusions about the orbital
contributions to pairing.
Based on the spectra of odd-even nuclei
for $  A_{c} \geq 64  $,
it is important to use a model space that includes
the $  0g_{9/2}  $ orbital to obtain the pairing
and deformed properties.
The computationaly demanding
configuration-mixing calculations for
$  A_{c} \geq 72  $ that include the "dip" at $  A_{c}=76  $
remain to be carried out.

{\bf Acknowledgements:}
BAB acknowledges support from NSF grant PHY-2110365.
The IMSRG calculations were performed with the codes imsrg$++$ \cite{imsrg}
and kshell \cite{kshell}.
The code imsrg$++$ utilizes the Armadillo C$++$
library \cite{armadillo1}, \cite{armadillo2}.

\end{document}